\documentclass[12pt]{article}

\usepackage{scicite}
\usepackage{graphicx}
\usepackage{times}
\usepackage{xcolor}
\usepackage{amsfonts}
\usepackage{esvect}
\usepackage{harpoon}
\usepackage{amsmath,accents}

\usepackage{subfigure,color}
\renewcommand{\figurename}{Fig.}

\def\e{\begin{equation}}
\def\f{\end{equation}}
\def\_#1{{\bf #1}}
\def\/#1{_{\rm #1}}
\def\.{\cdot}

\newenvironment{sciabstract}{%
\begin{quote} \bf}
{\end{quote}}

\title{Spin-dependent phenomena at chiral temporal interfaces}

\author
{M.~H.~Mostafa$^{\ast}$, M.~S.~Mirmoosa, S.~A.~Tretyakov$^{\ast}$
\\
\normalsize{Department~of~Electronics~and~Nanoengineering, Aalto~University, Espoo, Finland}\\
\normalsize{$^\ast$ Correspondence: mohamed.mostafa@aalto.fi, sergei.tretyakov@aalto.fi}
}

\date{}

\begin{document} 

\baselineskip24pt

\maketitle 

\begin{sciabstract}

{Temporally varying electromagnetic media have been extensively investigated recently to unveil new means for controlling light. However, spin-dependent phenomena in such media have not been explored thoroughly. Here, we reveal the existence of spin-dependent phenomena at a temporal interface between chiral and dielectric media. In particular, we show theoretically and numerically that due to the material discontinuity in time, linearly polarized light is split into forward-propagating right-handed and left-handed circularly polarized waves having {\it different angular frequencies} and {\it the same phase velocities}. This salient effect allows complete temporal separation of the two spin states of light with high efficiency. In addition, a phenomenon of spin-dependent gain/loss is observed. Furthermore, we show that when the dielectric medium is switched again to the original chiral medium, the right- and left-handed circularly polarized light waves (with different angular frequencies) merge to form a linearly polarized wave. Our findings extend spin-dependent interactions of light from space to space-time.}

\end{sciabstract}

\noindent 


\section*{Introduction}
\label{sec1}
Manipulating effective properties of electromagnetic systems in time provides an exceptional opportunity to control light and attain unique responses~\cite{review}. Recently, by exploiting this approach, a multitude of wave phenomena and applications have been uncovered, including nonreciprocity~\cite{Fang,bi,nonrec-rev,luminal}, frequency conversion~\cite{Lee, Liu,Shaltout}, time reversal~\cite{Bacot}, antireflection temporal coating~\cite{Pacheco}, and more~\cite{PendryRev, Sajjad-review}. However, spin-dependent interactions in time-varying media have not been contemplated thoroughly~\cite{Kalluri92, Kalluri93, Kalluri2010, Huanan}. In addition, most studies have mainly focused on temporal manipulations of isotropic or anisotropic systems~\cite{temporalaiming, Akbarzadeh}, while less attention is given to time-varying bianisotropic media or systems~\cite{bi}. On one hand, it has been shown that a traveling wave space–time modulation that emulates a moving media creates bianisotropic-like coupling~\cite{Huidobro,Huidobro2}. On the other hand, these works utilize temporal modulation to produce bianisotropic effects (magnetoelectric coupling), which is fundamentally different from manipulating a bianisotropic medium in time. From this point of view, temporal manipulations of bianisotropic media is an unexplored area of research while it holds potential to realize and use  spin-dependent interactions of a new type.

To introduce spin-dependent phenomena at temporal interfaces, it is helpful to refer to spin-orbit interactions of light~\cite{Fortuno,Aiello} which has become an extremely active topic during recent years~\cite{Shi, Eismann, Chen, Higo}. A time-harmonic electromagnetic wave can be fully described by its intensity, wavevector, polarization state, and angular frequency. The intensity and wavevector of a wave represent its spatial degrees of freedom, while the angular frequency represents the temporal degree of freedom. Spin-orbit interactions take place when the spatial (orbital) degrees of freedom depend on the polarization (spin) of the propagating wave. In other words, the polarization state of the electromagnetic wave defines how the wave propagates through \emph{space}. An example of spin-orbit interactions is the spin-Hall effect of light~\cite{spin,Bliokh}, where a transverse spin-dependent subwavelength shift takes place at a spatial planar interface. This effect is utilized to spatially split/decompose linearly polarized light into right-handed and left-handed circularly polarized (RHCP/LHCP) waves~\cite{Hosten,Zhou}. It appears possible to expect similar effects at bianisotropic temporal interfaces, where a spin-dependent frequency shift would take place. In this case, the polarization state of the electromagnetic wave defines how the wave propagates through \emph{time}. To investigate this possibility, we study abrupt changes of bianisotropic chiral media parameters as functions of time. One of the main motivations is a possibility to switch \emph{spatial} mirror-inversion symmetry by varying some material parameters \emph{in time}.

In this paper, we make an initial step in this direction and contemplate a nonstationary chiral medium. In particular, we consider a temporal interface between a chiral medium and a dielectric medium. It is useful to conceptualise the problem as a temporal interface between two symmetries. This is an interface between spatially mirror-inversion asymmetric and symmetric media, hence, the symmetry is switched in time. In a chiral medium, the RHCP and LHCP waves which form a linearly polarized propagating wave are associated with different phase velocities~\cite{Ari}, which is the main property of chiral media. This property is not exclusive to chiral media, it is associated also with nonreciprocal magnetized and magneto-optical materials. In chiral media, such difference in the phase velocity of eigenwaves arises from spatial dispersion in materials with broken mirror-inversion symmetry. We show that due to this important characteristic, at a temporal interface between chiral and dielectric media the angular frequencies of the propagating RHCP and LHCP waves (composing a linearly polarized wave) are shifted to two different angular frequencies resulting in splitting the polarization states temporally. In addition, the energy density of the RHCP and LHCP waves experiences a spin-dependent gain/loss effect. 
Such phenomena constitute examples of an unusual class of wave-matter interactions.

Moreover, if the dielectric medium is switched back to the original chiral medium, the decomposed RHCP and LHCP waves merge again and form a linearly polarized propagating wave. Hence, temporal discontinuities in bianisotropic chiral media can also be used to merge RHCP and LHCP waves (with different angular frequencies) to compose linearly polarized light. Finally, we calculate the amplitudes of forward and backward waves generated due to chiral-dielectric temporal discontinuities and prove that under certain conditions, the backward waves (reflected waves) vanish, meaning that only the forward waves are propagating after temporal discontinuities.  


\section*{Time-domain model of chiral media
}\label{sec2}

To study time-varying chiral media, time-domain constitutive relations are needed. However, the commonly used constitutive equations of chiral media (so called Post and Tellegen relations)~\cite{Ari} are applicable only in the frequency domain. This is due to the fact that electromagnetic chirality is a manifestation of spatial dispersion, leading to inevitable frequency dispersion of chirality parameters in both these models. For this reason, we use the Condon model, which connects the electric flux density to the time derivative of the magnetic field and the magnetic flux density to the time derivative of the electric field~\cite{CONDON}.
This model, introduced in 1937, approximately models chirality effects with a non-dispersive parameter $g$, which is a crucial feature. The model is applicable at frequencies well below all resonances of chiral molecules or inclusions, where the rotatory power linearly decreases to zero at the limit of zero frequency \cite{Serdyukov}.   

In chiral media, a linearly polarized plane wave can be expressed as a combination of RHCP and LHCP waves having the same angular frequency but propagating at different phase velocities~\cite{Ari}. Splitting the fields of a plane wave into RHCP and LHCP components, we write the constitutive relations of isotropic chiral media as
\begin{subequations}
\begin{align}
& \_D= \Big[\overbrace{\epsilon_{\rm eff} \_E^-+ g \frac{\partial \_H^-}{\partial t}}^{\_D^-}\Big]+\Big[\overbrace{\epsilon_{\rm eff} \_E^+ + g \frac{\partial \_H^+}{\partial t}}^{\_D^+}\Big] \label{eq:one},\\
& \_B= \Big[\overbrace{\mu_{\rm eff} \_H^- - g \frac{\partial \_E^-}{\partial t}}^{\_B^-}\Big]+\Big[\overbrace{\mu_{\rm eff} \_H^+ - g \frac{\partial \_E^+}{\partial t}}^{\_B^+}\Big],
\label{eq:two}
\end{align}
\end{subequations}
in which $\epsilon_{\rm eff}$, $\mu_{\rm eff}$, and $g$ are the non-dispersive effective permittivity, effective permeability, and the chirality parameter (or rotatory parameter as Condon called it). The $\pm$ superscripts mark the RHCP and LHCP wave components, respectively. We consider electric and magnetic fields of a linearly polarized plane wave propagating in the $x$-direction as $\_E^{\pm}= {E_0\over2} \Big[\_{\hat{y}}\mp j\_{\hat{z}}\Big]e^{j(\omega t-\beta^{\pm} x)}$ and $\_H^{\pm}=\frac{\_{\hat{x}} \times \_E^{\pm}} {\eta_{\rm eff}}$, where $\eta_{\rm eff}$ is the medium effective intrinsic impedance, and $E_0$ is the complex amplitude of the electric field. We use the electrical engineering convention for time-harmonic oscillations (i.e.,~$\exp{(j\omega t)}$). By substituting the fields into Eqs.~\eqref{eq:one} and~\eqref{eq:two}, we arrive to the wavefield decomposition (see Supplementary Note 1)
\begin{equation}
\_D^{\pm}=\epsilon_{\rm eff}\Big(1 \mp \Psi \Big) \_E^{\pm},\,\,\,\
\_B^{\pm}= \mu_{\rm eff} \Big(1 \mp \Psi \Big) \_H^{\pm},
\label{eq:five}
\end{equation} 
in which $\Psi= g\omega c$, $\omega$ is the angular frequency, and $c=\frac{1}{{\sqrt{\epsilon_{\rm eff} \mu_{\rm eff}}}}$. The wavenumbers of plane waves in the two equivalent isotropic media equal to $\beta^{\pm}=\omega \sqrt{\epsilon_{\rm eff} \mu_{\rm eff}} (1 \mp \Psi)$~\cite{Ari}.

\section*{Temporal interface between chiral and dielectric media} 

\begin{figure}[t!]
\centerline
{\includegraphics[width=1\linewidth]{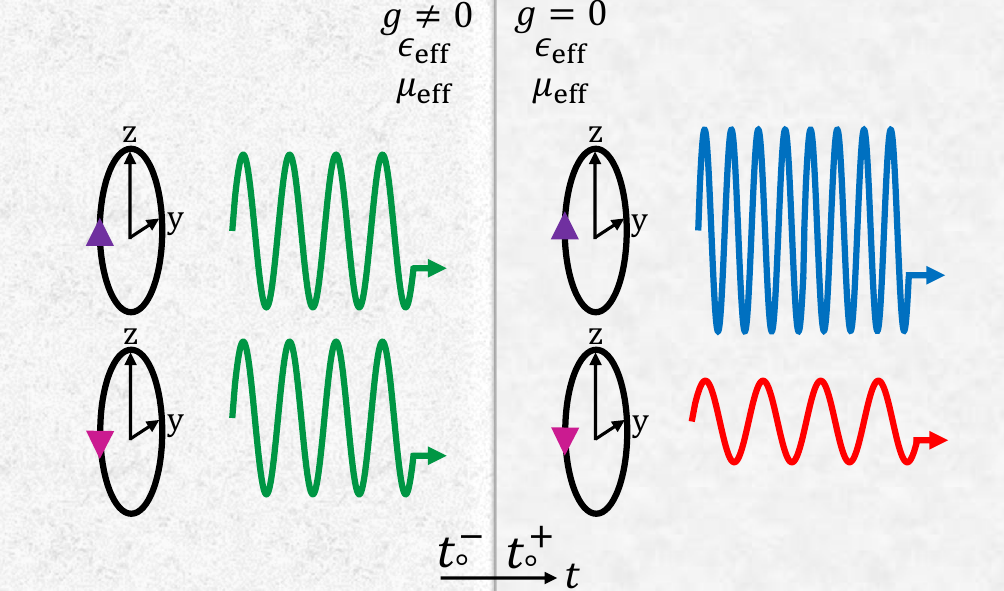} \label{1a}}
\caption{Schematic representation of a temporal interface between chiral and dielectric media. Spin-dependent frequency shift and spin-dependent gain/loss take place at the temporal interface.} 
\label{fig}
\end{figure} 

Here, we consider a temporal interface between isotropic chiral and dielectric media having the same effective permittivity and permeability (Fig.~\ref{fig}). 
We note that it is possible to solve the problem in the general case when permittivity and permeability experience jumps, arriving to similar physical results. We contemplate a chiral medium supporting a linearly polarized plane wave at frequency $\omega_1$ when the chirality parameter $g$ rapidly changes to zero, that is, the medium becomes nonchiral. In his paper published in 1958, Morgenthaler showed that the electric and magnetic flux densities are continuous at a temporal interface~\cite{Morgen}. Using that property, we write that $\_D^{\pm}_1= \_D^{\pm}_2$ and $\_B^{\pm}_1= \_B^{\pm}_2$, where the subscripts $1,2$ correspond to the fields before $(t=t^-_{0})$ and after $(t=t^+_{0})$ the temporal discontinuity, respectively ($t_{0}$ is the switching moment). According to Morgenthaler, after the temporal jump, there are forward and backward waves in analogy with a spatial interface at which we have transmitted and reflected waves. Keeping this in mind, we substitute Eq.~\eqref{eq:five} into $\_D^{\pm}_1= \_D^{\pm}_2$ and $\_B^{\pm}_1= \_B^{\pm}_2$. Assuming $t_0=0$ and after some mathematical manipulations, we find that (see Supplementary Note 2)
\begin{equation}
\epsilon_{\rm eff}(1\mp \Psi_1)\_E^{\pm}_1=\epsilon_{\rm eff}\overbrace{(\Upsilon_{\rm c\|d}^{\pm} + \Gamma^{\pm}_{c\|d})\_E^{\pm}_1}^{\_E^{\pm}_2},\,\,\,\,\,\,\,\,\mu_{\rm eff}(1\mp \Psi_1)\_H^{\pm}_1=\mu_{\rm eff}\overbrace{(\Upsilon_{\rm c\|d}^{\pm} - \Gamma^{\pm}_{c\|d})\_H^{\pm}_1}^{\_H^{\pm}_2},
\label{eq:nine}
\end{equation}
in which $\Psi_1=\omega_1 g c$ depends on the angular frequency before the temporal discontinuity. The forward and backward propagation coefficients (for a temporal interface between chiral and dielectric media) are denoted as $\Upsilon_{\rm c\|d}$ and $\Gamma_{\rm c\|d}$, respectively. The above expressions indicate that the polarization states and the phase constants are conserved at the temporal interface, as shown in Supplementary Note 2. Due to the conservation of the phase constant and by knowing that $\beta^{\pm}_1=\omega_1 \sqrt{\mu_{\rm eff}\epsilon_{\rm eff}} (1 \mp \Psi_1)$ and $\beta^{\pm}_2=\omega^{\pm}_2 \sqrt{\mu_{\rm eff}\epsilon_{\rm eff}}$, we arrive to the following important relation:
\begin{equation}
\omega^{\pm}_2=\omega_1\Big(1\mp\Psi_1\Big).
\label{eq:thirteen}
\end{equation}
This result shows that the RHCP and LHCP components have different angular frequencies after the temporal jump, meaning that the polarization states are separated temporally. In addition, from Eq.~\eqref{eq:nine} we find that $\Gamma^{\pm}_{c\|d}=0$ and $\Upsilon^{\pm}_{c\|d}=1\mp \Psi_1$. In other words, at fast transitions (transition time smaller than the wave period) from  a chiral medium to a dielectric one (while keeping the effective permittivity and permeability the same), no backward waves are generated.

The same result can be obtained using the Morgenthaler equations for forward and backward propagating waves at a temporal interface in dielectric media~\cite{Morgen}. These equations relate the forward and backward propagation coefficients to the permittivity and permeability of the medium before and after the temporal jump. Assume that the equivalent permittivity and permeability of the chiral medium for RHCP and LHCP equal $\epsilon_{\rm{eq}}^{\pm}=\epsilon_{\rm eff}\Big(1 \mp \Psi_1\Big)$ and $\mu_{\rm{eq}}^{\pm}=\mu_{\rm eff}\Big(1 \mp \Psi_1\Big)$, respectively. In this case, changing the chirality parameter in time is equivalent to changing the permittivities and permeabilities of two equivalent magnetodielectric media. Then, if we plug these parameters in the Morgenthaler equations as $\Upsilon^{\pm}_{c\|d}=\frac{1}{2}[\frac{\epsilon_{\rm{eq}}^{\pm}}{\epsilon_{\rm eff}}+\frac{\sqrt{\epsilon_{\rm{eq}}^{\pm} \mu_{\rm{eq}}^{\pm}}}{\sqrt{\epsilon_{\rm eff} \mu_{\rm eff}}}]$ and $\Gamma^{\pm}_{c\|d}=\frac{1}{2}[\frac{\epsilon_{\rm{eq}}^{\pm}}{\epsilon_{\rm eff}}-\frac{\sqrt{\epsilon_{\rm{eq}}^{\pm} \mu_{\rm{eq}}^{\pm}}}{\sqrt{\epsilon_{\rm eff} \mu_{\rm eff}}}]$, we get the same $\Upsilon_{\rm c\|d}$ and $\Gamma_{\rm c\|d}$ as the ones derived above. From the expressions of $\epsilon_{\rm{eq}}^{\pm}$ and $\mu_{\rm{eq}}^{\pm}$, it is evident that no waves propagate backward, as the equivalent wave impedance is the same in both media~\cite{Morgen} and equals $\sqrt{\frac{\mu_{\rm eff}}{\epsilon_{\rm eff}}}$. On the other hand, if the permittivity of the background medium is also switched together with the chirality parameter, a backward wave will be created due to a jump in the equivalent wave impedance. 

Moreover, as the RHCP and LHCP waves propagate in media having different equivalent permittivities and permeabilities before the temporal interface, a wave having one of the polarization states exhibits loss and the other one exhibits gain, introducing spin-dependent gain/loss. The energy density gain/loss (defined as the ratio of the energy density before and after the temporal interface) is equal to $\frac{1}{2}(\frac{\epsilon_{\rm{eq}}^{\pm}}{\epsilon_{\rm eff}}+\frac{\mu_{\rm{eq}}^{\pm}}{\mu_{\rm eff}})$ \cite{Morgen}, which simplifies to $1\mp \Psi_1$.


To better understand how spin of light interacts with bianisotropic temporal discontinuities, it is important to investigate what happens if the dielectric medium (after the temporal interface) is at a later moment of time switched back to the same chiral medium, representing a ``temporal slab.'' In this case the chiral medium is considered twice. For both cases the fields in the chiral medium are defined using the same parameters,  and subscripts 1 and 3 are used to distinguish between them. Following the same analysis method as before, we find that the forward and backward propagation coefficients for the second temporal interface are $\Upsilon^{\pm}_{d\|c}=\frac{1}{1\mp \Psi_3^{\pm}}$ and $\Gamma^{\pm}_{d\|c}=0$, respectively, where $\Psi_3^{\pm}=\omega_3^{\pm} g c$. Again,  no backward propagating waves are created. Interestingly, the transmission coefficient is a function of the angular frequency after the second temporal interface. This angular frequency can be calculated from the conservation of wavenumbers $\beta^{\pm}_2=\omega^{\pm}_2 \sqrt{\mu_{\rm eff}\epsilon_{\rm eff}}$ and $\beta^{\pm}_3=\omega_3^{\pm} \sqrt{\mu_{\rm eff}\epsilon_{\rm eff}} (1 \mp \Psi_3^{\pm})$. As $\beta^{\pm}_3$ is a quadratic equation for $\omega_3^{\pm}$, equations $\beta^{\pm}_2=\beta^{\pm}_3$ have two solutions for frequencies $\omega_3^{\pm}$ after the second temporal interface, which read $\omega_3^{\pm}=\omega_1$ and $\omega_3^{\pm}=\pm\frac{ 1}{cg}-\omega_1$. The first solution indicates that the created RHCP and LHCP waves have the same frequency as the initial wave before the first interface. Since $\omega_3^{\pm}=\omega_1$, the total  forward propagation coefficient  $\Upsilon^{\pm}_{d\|c}\cdot \Upsilon^{\pm}_{c\|d}=1$, meaning that the amplitude of the created wave is equal to the initial amplitude. Thus, the initial linearly polarized wave is formed again according to the first solution.
We note that this phenomenon can be used to merge two differently polarized waves (at different angular frequencies) into a linearly polarized wave. On the other hand, the second solution $\omega_3^{\pm}=\pm\frac{ 1}{cg}-\omega_1$ 
indicates that, after the second interface, additional waves can be created at high frequencies. The Condon model of chiral media is applicable only at enough low frequencies, well below the resonant frequencies of chiral particles forming the medium. However, for realistic values of $g$ and considering relatively small $\omega_1$ that is far from resonance, the term $\frac{1}{cg}$ is much larger than $\omega_1$, leading to very large $\omega_3$ (see Supplementary Note 2). Thus, the Condon model cannot be reliably used to get accurate results for the second solution. More broadband time-domain models of chiral media need to be used to investigate the second solution, and for this reason we do not consider this solution here.


\section*{Realization and numerical results} 

As a specific realization, we consider canonical metal-wire chiral particles, where the particles are formed of two short straight wires (arm length $l$) connected to an electrically small loop (the loop area $S$) \cite{Serdyukov}. The electromagnetic fields defined above have transversal components in the $\_{\hat{y}}\rm{-}\_{\hat{z}}$ plane, thus, we consider an uniaxial chiral medium composed of two orthogonal arrays of small chiral particles (Fig.~2A). All the medium parameters are expressible in dyadic form as $\overline{\overline{a}}=a~ \overline{\overline{I_t}}+b~\_{\hat{x}}\_{\hat{x}}$, where $a$ and $b$ are scalars or pseudoscalars, $\_{\hat{x}}$ is the unit vector along the preferred direction in the medium, and $\overline{\overline{I_t}}$ is the transverse unit dyadic. As there is magnetoelectric coupling only in the transverse plane, switching chirality induces temporal interface for transversal fields only. Hence, to the end of the paper, we solve the problem in the transverse plane, which reduces the mathematical representation. In this case, the chirality parameter dyadic $\overline{\overline{g}}=g\overline{\overline{I_t}}$ reduces to pseudoscalar $g$, similar to all the other parameter dyadics. 

\begin{figure}[]
\centering
\subfigure{\includegraphics[width=0.555\linewidth]{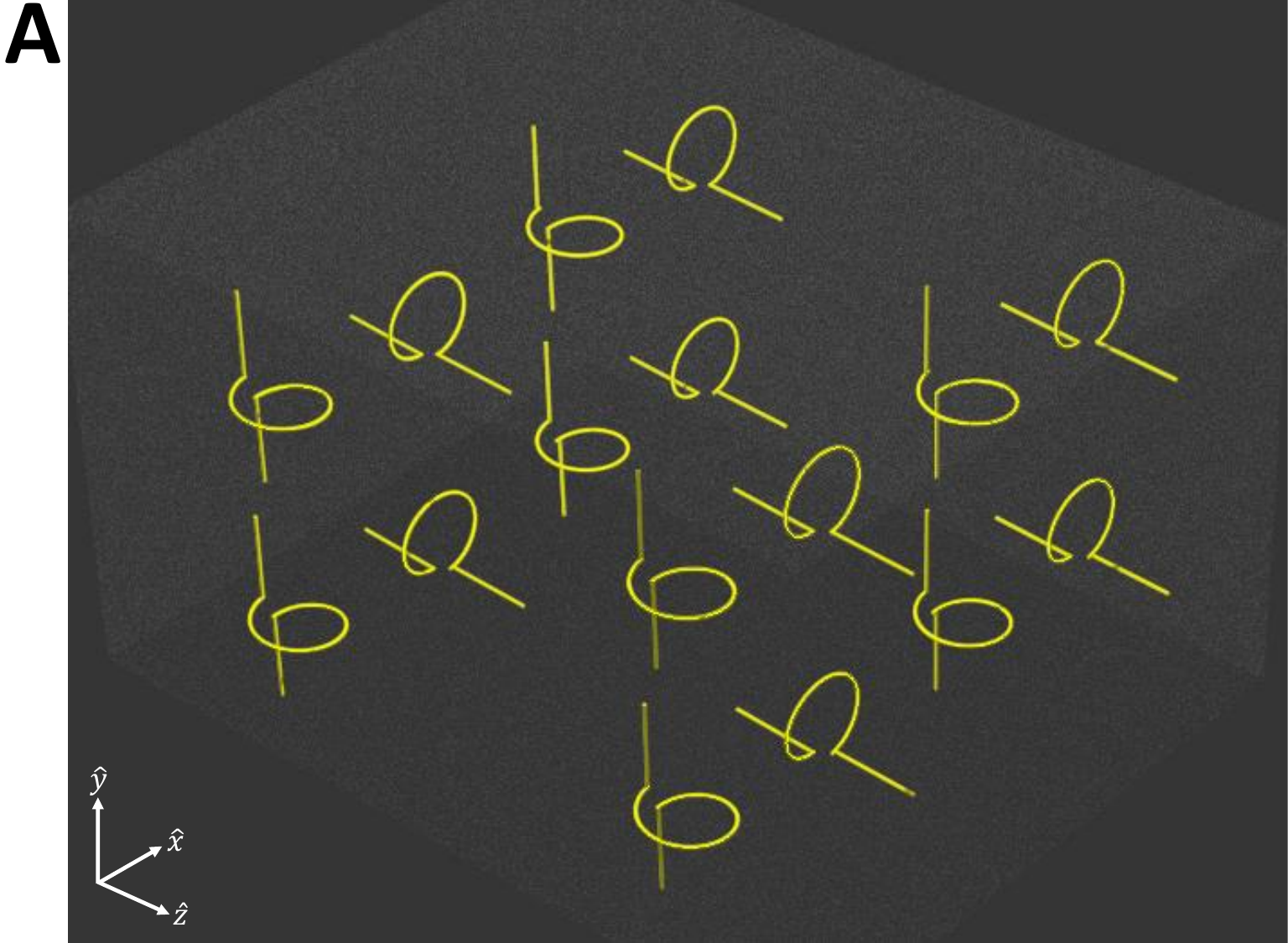}\label{2a}}
\subfigure{\includegraphics[width=0.45\linewidth]{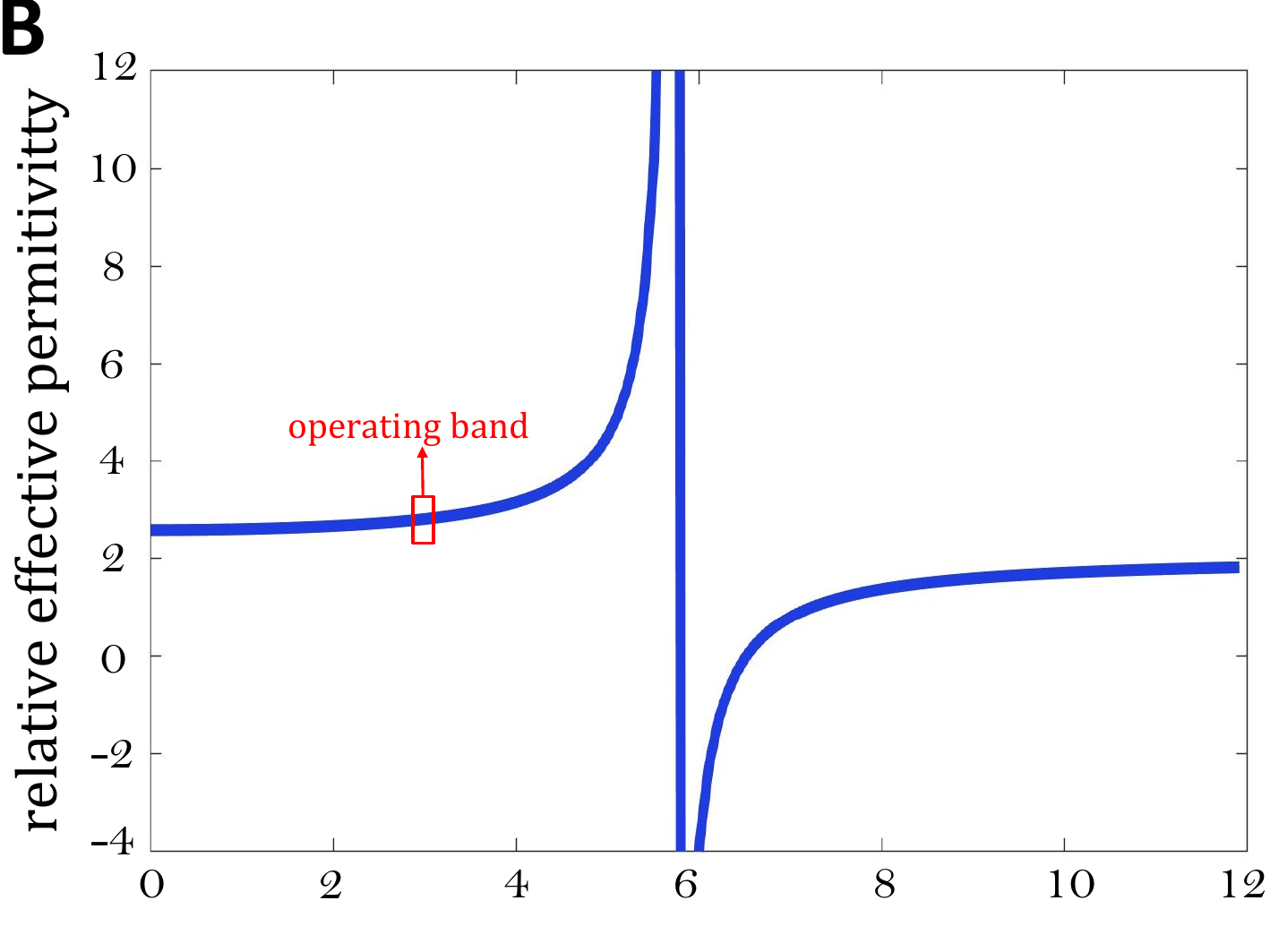}\label{2b}}
\subfigure{\includegraphics[width=0.45\linewidth]{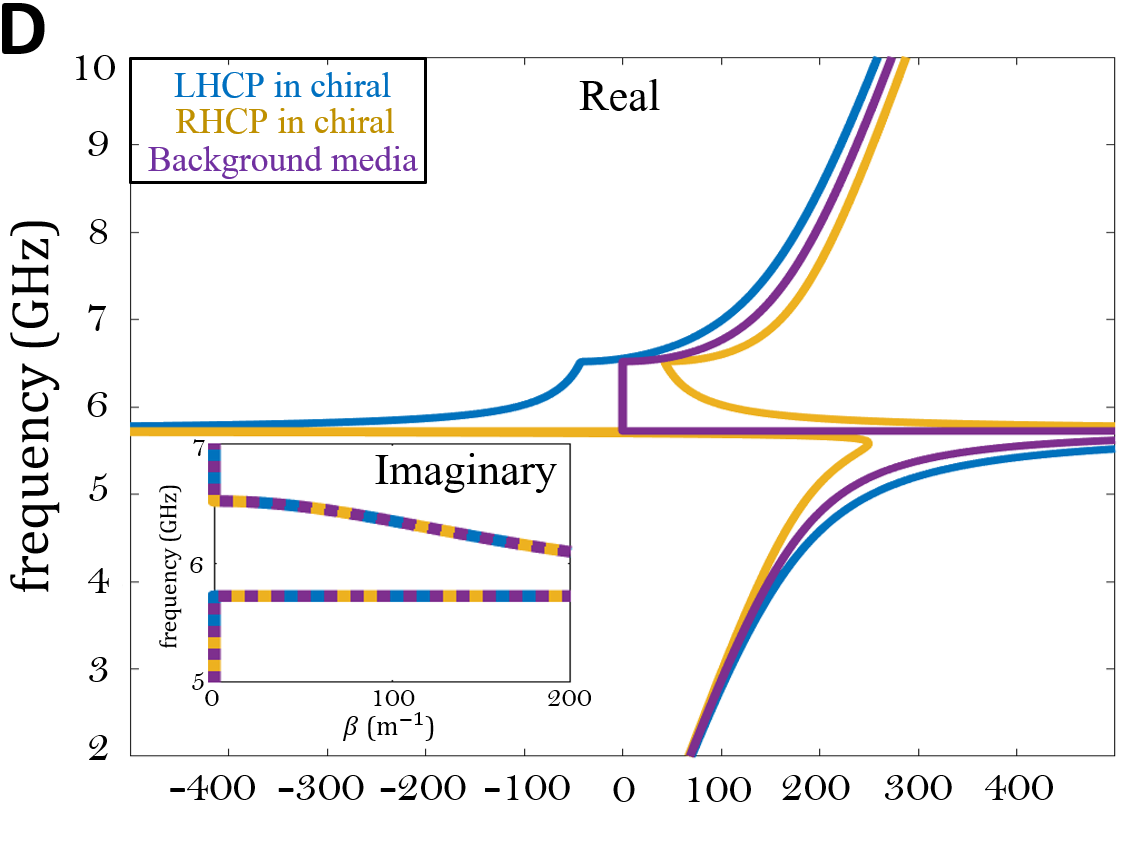}\label{2d}}
\subfigure{\includegraphics[width=0.45\linewidth]{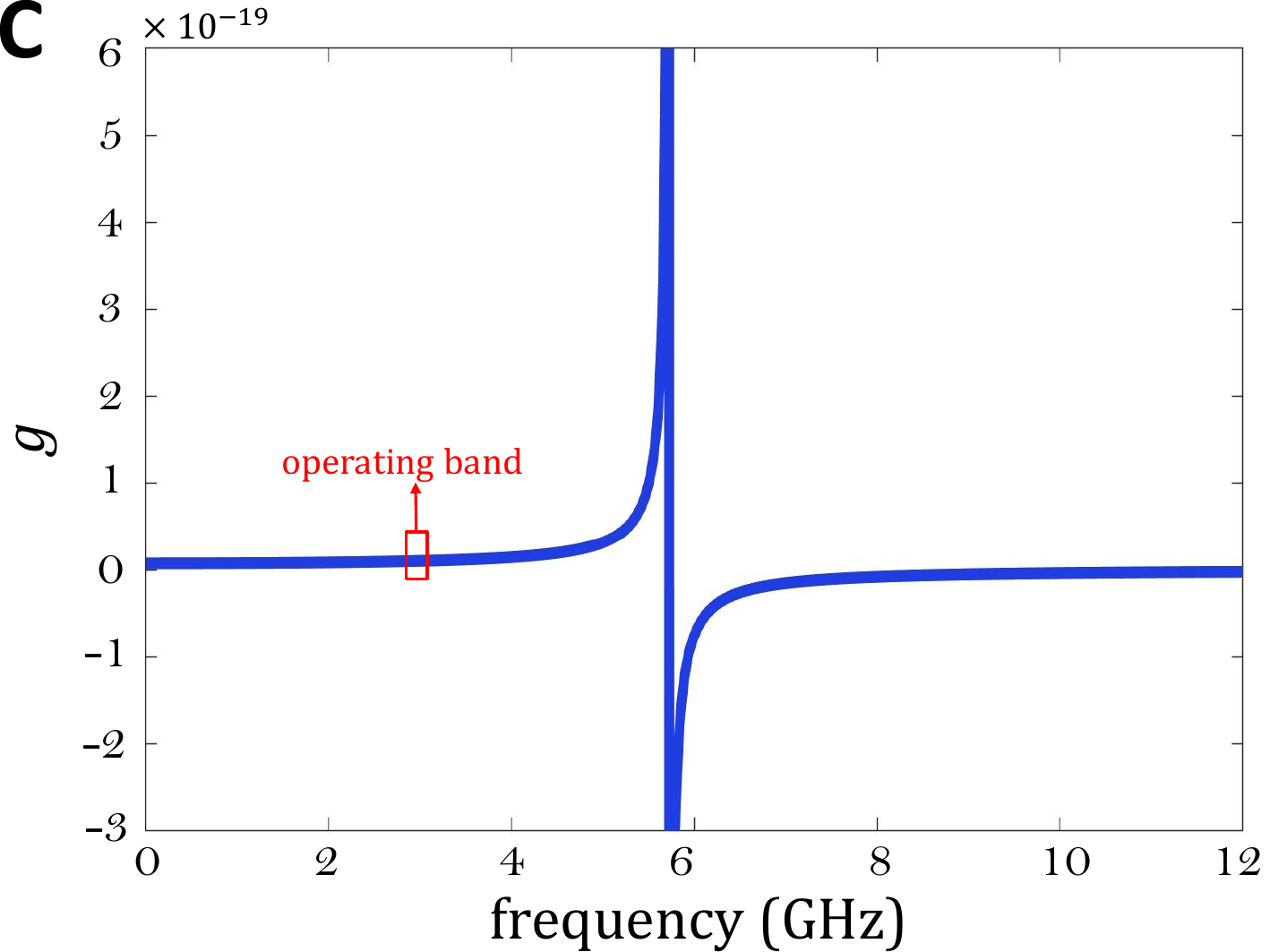}\label{2c}}
\subfigure{\includegraphics[width=0.45\linewidth]{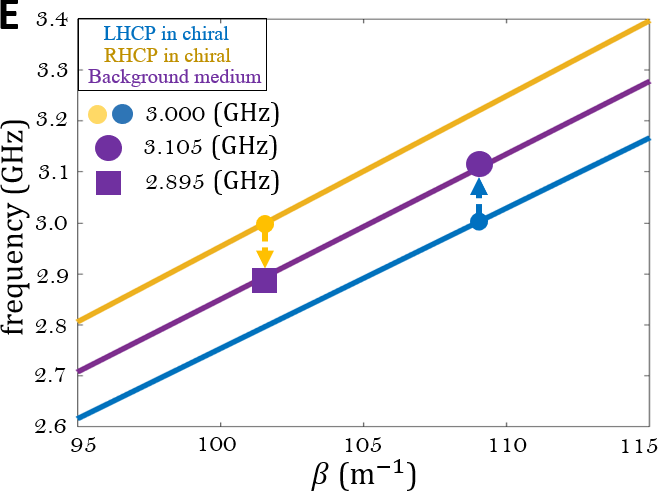}\label{2e}}
\caption{ (\textbf{A}) Uniaxial chiral composite having magnetoelectric coupling in the $\_{\hat{y}} \rm{-} \_{\hat{z}}$ plane. (\textbf{B}) Relative effective permittivity as a function of frequency. (\textbf{C}) The chirality parameter $g$ as a function of frequency. (\textbf{D}) Dispersion curves corresponding to RHCP and LHCP plane waves propagating in chiral medium, and, also, the dispersion curve associated with the dielectric medium. (\textbf{E}) Dispersion curve for the case when the effective permittivity and chirality parameter are fixed values and do not depend on the frequency. The arrows illustrate the frequency conversion phenomenon.} 
\label{conditions}
\end{figure} 

In the frequency domain, the effective material relations for chiral media are written as
$\big(\begin{smallmatrix}
  \_D \\
  \_B 
\end{smallmatrix}\big)$ 
$=$
$\big(\begin{smallmatrix}
  \epsilon_{\rm eff} & \alpha_{\rm eff}\\
  -\alpha_{\rm eff} & \mu_{\rm eff}
\end{smallmatrix}\big)$ 
$\cdot$
$\big(\begin{smallmatrix}
  \_E \\
  \_H 
\end{smallmatrix}\big)$, where $\alpha_{\rm eff}$ is the effective chirality parameter. The uniaxial particle polarizabilities can be estimated as $\alpha_{\rm ee}=\frac{1}{2}\frac{l^2 C}{1-\omega^2 LC}$, $\alpha_{\rm mm}=\frac{1}{2} \frac{\mu^2 \omega^2 S^2 C}{1-\omega^2 LC}$, and $\alpha_{\rm em}=-\alpha_{\rm me}=\frac{1}{2}\frac{j\omega \mu S l C}{1-\omega^2 LC}$ (see Supplementary Note 3), where $\mu$ is the permeability of the background medium, $C$ and $L$ are the capacitance and inductance of the wire antenna and the loop respectively. To estimate the effective material parameters of the composite, we use the Maxwell-Garnett model for mixtures of bianisotropic particles \cite{JEWA,Serdyukov}. Well below the particle resonance, the magnetic polarizability is negligible, being a second-order spatial dispersion effect, and a non-magnetic background medium is considered, thus, we have $\mu_{\rm eff}=\mu_0$. Enough small number of particles per unit volume $N$ is selected, so that switching chirality would not affect the effective permittivity of the composite. This property is ensured if ${N^2 \alpha_{\rm em}^2 \over  9\epsilon\mu}\ll 1$, where $\epsilon$ is permittivity of the background medium (see Supplementary Note 3). In this case, the effective permittivity and effective chirality parameter can be estimated by the classical Maxwell-Garnett formula for electrically-polarizable particles as $\epsilon_{\rm eff}=\epsilon_0+{N\alpha_{\rm ee}\over 1-{N\alpha_{\rm ee}\over 3\epsilon}}$ and $\alpha_{\rm eff}={-N\alpha_{\rm me}\over 1-{N\alpha_{\rm ee}\over 3\epsilon}}$, respectively (see Supplementary Note 3). To this point, the effective parameters are dispersive and resonant due to the denominator $1-\omega^2LC$. We design the chiral composite and choose the operating frequency such that we operate reasonably far from the  resonance at all moments, before and after the temporal interface. Thus, $1-\omega^2LC$ is approximately constant at all moments of time, which is a condition that has to be met while designing the chiral composite obeying the Condon model. Keeping in mind that $1-\omega^2LC$ is constant, the non-dispersive chirality parameter $g$ can be written in terms of the effective chirality parameter as $g=\frac{\alpha_{\rm{eff}}}{j\omega}$. Transforming the effective chirality parameter to time domain leads to $\alpha_{\rm{eff}}=g \frac{\partial }{\partial t}$. The non-dispersive time-domain material relations can be written as 
$\big(\begin{smallmatrix}
  \_D \\
  \_B 
\end{smallmatrix}\big)$ 
$=$
$\big(\begin{smallmatrix}
  \epsilon_{\rm eff} & g \frac{\partial }{\partial t}\\
  -g \frac{\partial }{\partial t} & \mu_{\rm eff}
\end{smallmatrix}\big)$ 
$\cdot$
$\big(\begin{smallmatrix}
  \_E \\
  \_H 
\end{smallmatrix}\big)$, which is the Condon model. Finally, to modulate $g$ without affecting $\epsilon_{\rm eff}$ or $\mu_{\rm eff}$, we modulate the loop area $S$. In practical realizations, chiral mixtures can be transformed to racemic ones using switches at the connection of the loops to the straight wires, to reverse handedness of one half of the particles. 

Consider the fields $\_E^{\pm}$ and $\_H^{\pm}$, which represent RHCP and LHCP plane waves that constitute a linearly polarized wave propagating in a chiral medium. We arbitrarily set $E_0=2$~V/m. While designing the chiral medium parameters, there are two conditions that have to be satisfied. Firstly, $\Psi_{1}$ should have a considerably large value to induce a considerable frequency shift, as the spin-dependent frequency shift is a function of $\Psi_{1}$. Secondly, the condition of operating reasonably far from the resonance should be met to keep $1-\omega^2LC$ constant. Taking these conditions in consideration, the arm length $11$~mm, loop radius $2$~mm, frequency $\omega_1/(2 \pi)=3$~GHz, background relative permittivity $2$, and volume fraction $0.15$ have been chosen, corresponding  to $g=1.0348 \times 10^{-20}~\rm s^2/m$, $\Psi_{1}=0.035$ and a considerable frequency shift of $105$~MHz. Figures~2B and~2C show how $\epsilon_{\rm eff}$ and $g$ depend on the frequency. It can be seen that the two conditions are inversely proportional, that is, increasing $\Psi_{1}$ requires getting closer to the resonance. Hence, the frequency shift that can be achieved is limited if we stay within the applicability of the Condon model. However, generally, the spin-dependent frequency shift is not necessarily limited, it is limited in our case because of the limitations of the used models. 
The effective parameters are approximately constant within the operating band that is defined as $3 \pm 0.105$~GHz (Figs.~2B and 2C). To show it explicitly, Fig.~2D presents the dispersion curves for the RHCP and LHCP plane waves. Indeed, within this narrow frequency range whose center is at 3 GHz the dispersion curve is nearly linear, similar to a tilted light line. Thus, both conditions are satisfied. Consequently, we have $\mu_{\rm eff}=\mu_0 \ \rm H/m$ and $\epsilon_{\rm eff}= 2.49 \times 10^{-11} ~\rm F/m$. Accordingly, the equivalent parameters for the wavefields in this chiral medium read  $\epsilon_{\rm{eq}}^{\pm}=\epsilon_{\rm eff}\Big(1 \mp \Psi_{1}\Big)$ and $\mu_{\rm{eq}}^{\pm}=\mu_{\rm eff}\Big(1 \mp \Psi_{1}\Big)$. To finalize the design, we check that the condition of negligible effect of varying chirality on the effective permittivity is satisfied. Substituting the above values, we find that ${N^2 \alpha_{\rm em}^2 \over 9\epsilon\mu}\approx 0.0001\ll 1$. Thus, all the validity conditions for used models are satisfied. 

We verify the presented theory numerically using the time-domain solver of the commercial software COMSOL Multiphysics®. In simulations we study two temporal discontinuities at $t_1$ and $t_2$. At the first discontinuity at $t_1$, the medium properties change in time from those of this chiral medium to a simple dielectric medium having $\mu_{\rm eff}=\mu_0$, $\epsilon_{\rm eff}= 2.49 \times 10^{-11} ~\rm F/m $, and $g=0$, meaning that the medium has the same effective permittivity and permeability before and after the time discontinuity. At the second discontinuity at $t_2$, the medium properties change again in time to the same chiral medium with $g=1.03 \times 10^{-20} ~\rm s^2/m $. The transition time is equal to $\frac{1}{3}$ of the wave period. According to the  theoretical results presented above, there should be no generated  backward waves ($\Gamma^{\pm}_{c\|d},\Gamma^{\pm}_{d\|c}=0$), and the forward propagating waves should have the transmission coefficients and angular frequencies given by $\Upsilon^{\pm}_{c\|d}=1\mp 0.035$ and ${\omega^{\pm}_2\over 2 \pi}=  3(1\mp0.035)~\rm GHz $ after the first discontinuity, as shown by Fig.~2E, while after the second discontinuity we should get $\Upsilon^{\pm}_{c\|d}.\Upsilon^{\pm}_{d\|c}=1$ and ${\omega^{\pm}_3 /(2 \pi)}= 3~\rm GHz$.

\begin{figure}[t!]
\centering
\subfigure{\includegraphics[width=0.90\linewidth]{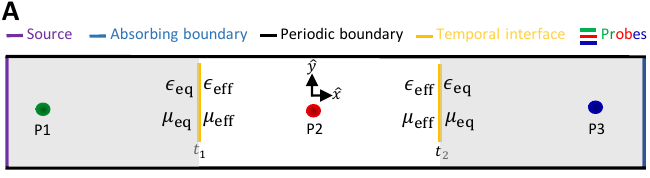}\label{setup}}
\hspace{2em}%
\subfigure{\includegraphics[width=0.49\linewidth]{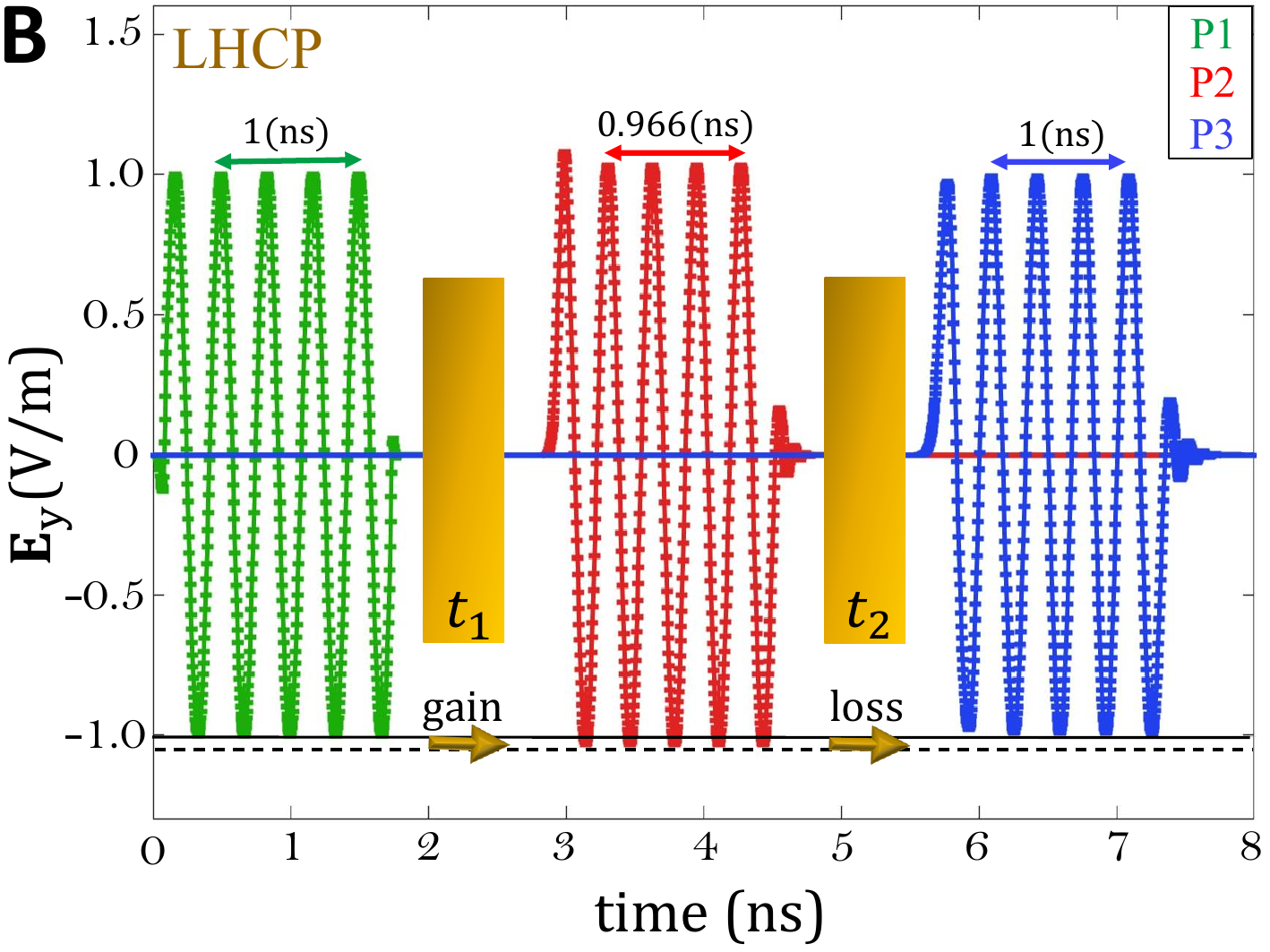}\label{3b}}
\subfigure{\includegraphics[width=0.49\linewidth]{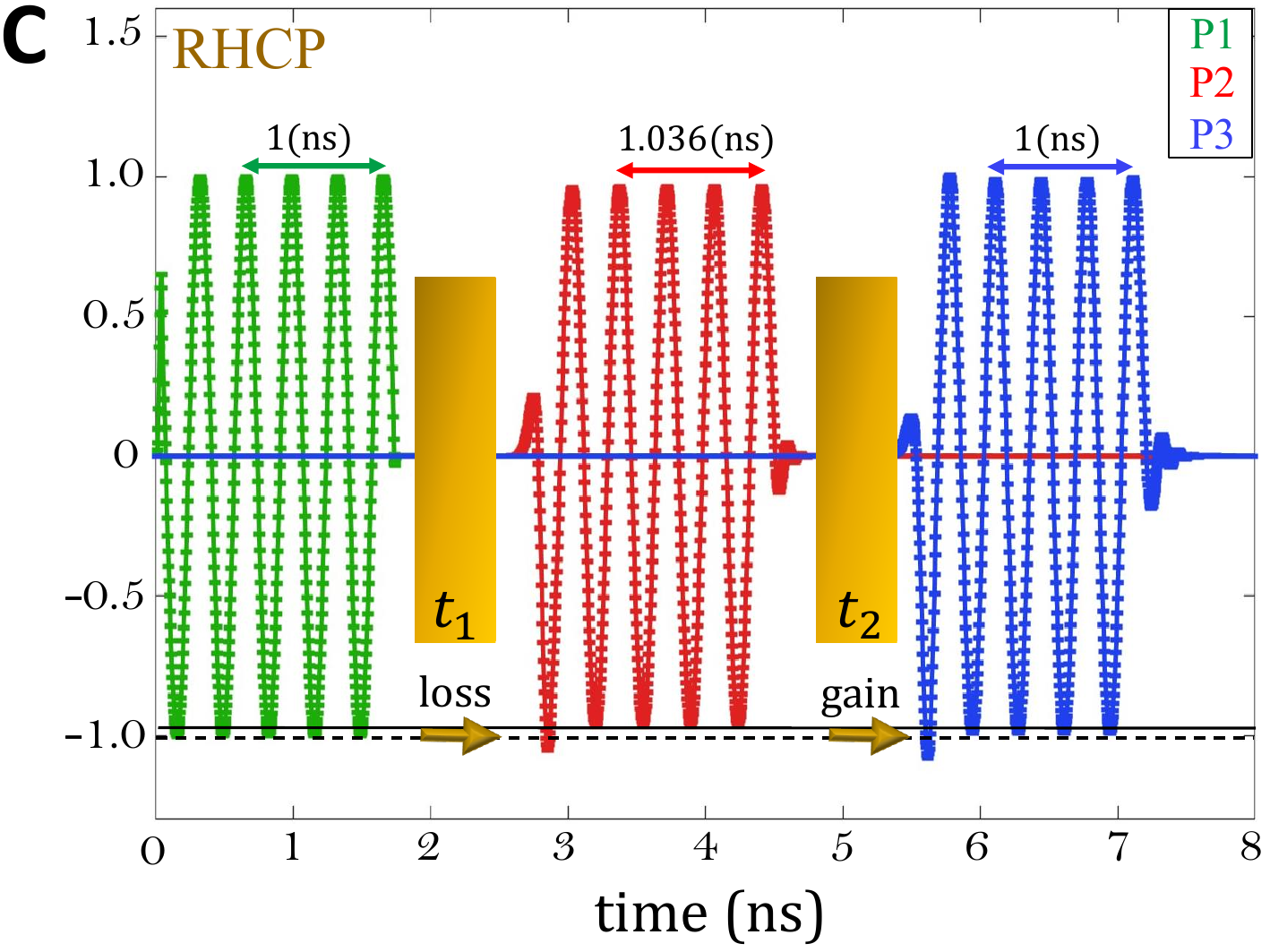}\label{3c}}
\caption{Numerical simulation of temporal interfaces between chiral and dielectric media using COMSOL Multiphysics. (\textbf{A}) Simulation domain. The chiral medium is associated with equivalent permittivity $\epsilon_{\rm{eq}}^{\pm}$ and equivalent permeability $\mu_{\rm{eq}}^{\pm}$, while the dielectric medium is associated with permittivity $\epsilon_{\rm eff}$ and permeability $\mu_{\rm eff}$. Three probes are used to measure the fields. Before the first temporal interface (before $t_1$), probe~1 measures the incident fields. On the other hand, after the first temporal interface (after $t_1$ and before $t_2$) probe~1 measures the backward propagating waves, and probe~2 measures the forward propagating waves. Similarly, after the second temporal interface (after $t_2$) probe~2 measures the backward propagating waves, and probe~3 measures the forward propagating waves. (\textbf{B}) $\_E_y(t)$ of the LHCP wave before and after the temporal interfaces. (\textbf{C}) $\_E_y(t)$ of the RHCP wave before and after the temporal interfaces.} 
\label{num_sim}
\end{figure} 

To simplify the data analysis, we simulate the RHCP and LHCP waves separately. The simulation domain is shown in Fig.~3A.
Figures~3B and~3C show the $y$-component of electric field for the incident, backward, and forward propagating  waves. The amplitudes and frequencies are in agreement with the theoretical predictions given above. Similar results are obtained for the $z$-component of the electric field and for the magnetic field components (see Supplementary Note 4). It can be seen that there are no backward propagating waves, while the amplitudes and frequencies for RHCP and LHCP forward propagating waves are different after the first temporal interface. Hence, the first temporal interface efficiently splits the polarization states of the incident wave, as the RHCP and LHCP waves are propagating at considerably different angular frequencies, due to a frequency shift of approximately $103$ and $107$~MHz for the LHCP and RHCP waves, respectively (theory predicts 105~MHz). In addition, the simulation also confirms that there is a spin-dependent gain/loss, as the amplitudes change by approximately $0.031$ and $0.039$~V/m for the LHCP and RHCP waves, respectively (theory predicts 0.035~V/m). There are some insignificant differences between the theoretical and numerical results, which is expected due to considering discontinuous waves in simulations and also changing the material parameters smoothly.
After the second temporal jump the RHCP and LHCP propagating waves combine again to constitute a linearly polarized wave, as both RHCP and LHCP propagate again at the same angular frequency. These numerical results confirm that temporal manipulations of bianisotropic chiral media control the spin of light and induce a spin-dependent frequency shift and spin-dependent gain/loss.


\section*{Discussion and outlook}\label{sec13}

This work contemplated a nonstationary chiral medium and unveiled spin-dependant phenomena at chiral temporal interfaces. Specifically, we showed that due to abruptly removing/inducing mirror-inversion asymmetry, a spin-dependent phenomena take place resulting in splitting/merging the spin states of light. Furthermore, we showed that temporal discontinuity of chiral media induce spin-dependent energy density gain/loss. Our results leverage time as one more degree of freedom for spin-controlled manipulations of light, which can lead to applications in electromagnetics, photonics, and quantum information processing. Potential applications include, for example,  sensing and separating chiral particles and molecules in chemical production.

Our results serve as an initial study of spin-dependant phenomena at temporal interfaces, paving the way to many opportunities. Presenting a general theory that takes into consideration the temporal dispersion of nonstationary chirality is an important direction in the future. In addition, scrutinizing periodic modulations of chirality and chiral time crystals is certainly promising and intriguing. Also, expanding this work to other bianisotropic media (like omega medium) appears inevitable. Finally, experimental realizations are crucial. one proposal is to use canonical chiral particles made of thin conducting wires and utilize switches that connect the dipole arms to the loop. The two positions of switches would correspond to the opposite signs of the chirality parameter of each meta-atom. Thus, switching one half of the particles to the opposite state, one transforms a chiral composite to a non-chiral state without affecting the effective permittivity and permeability. Beside this proposal, spin-dependent phenomena are not exclusive to nonstationary chirality, as nonstationary nonreciprocal magnetized and magneto-optical media should exhibit spin-dependent interactions too. Consequently, in addition to the system proposed in this article, experimental verification is also feasible using temporally switched magnetization.


\bibliography{sn-bibliography}

\begin{thebibliography}{9}



\bibitem{review} 
A. M. Shaltout, V. M. Shalaev, M. L. Brongersma, Spatiotemporal light control with active metasurfaces. {\itshape Science} \textbf{364}, 6441 (2019).

\bibitem{Fang} 
K. Fang, J. Luo, A. Metelmann \textit{et al.}, Generalized non-reciprocity in an optomechanical circuit via synthetic magnetism \& reservoir engineering. {\itshape Nature Physics} \textbf{13}, 465–471 (2017).

\bibitem{bi}
X. Wang, G. Ptitcyn, V. S. Asadchy, A. D\'{\i}az-Rubio, M. S. Mirmoosa, F. Shanhui, S.A. Tretyakov, Nonreciprocity in bianisotropic systems with uniform time modulation. {\itshape Phys. Rev. Lett.} \textbf{125}, 26 (2020).

\bibitem{nonrec-rev}
D. L. Sounas, A. Alù, Non-reciprocal photonics based on time modulation. {\itshape Nature Photonics} \textbf{11}, 774–783 (2017).

\bibitem{luminal}
E. Galiffi, P. Huidobro, J. Pendry, Broadband nonreciprocal amplification in luminal metamaterials. {\itshape Phys. Rev. Lett.} \textbf{123}, 206101 (2019).

\bibitem{Lee}
K. Lee, J. Son, J. Park \textit{et al.}, Linear frequency conversion via sudden merging of meta-atoms in time-variant metasurfaces. {\itshape Nature Photonics} \textbf{12}, 765–773 (2018).

\bibitem{Liu}
M. Liu, D. Powell, Y. Zarate, I. Shadrivov, Huygens’ metadevices for parametric waves. {\itshape Phys. Rev. X} \textbf{8}, 031077 (2018).

\bibitem{Shaltout}
A. M. Shaltout \textit{et al.}, Spatiotemporal light control with frequency gradient metasurfaces. {\itshape Science} \textbf{365}(6451), 374–377 (2019).

\bibitem{Bacot} 
A. M. Bacot, M. Labousse, M. Eddi \textit{et al.}, Time reversal \& holography with spacetime transformations. {\itshape Nature Physics} \textbf{12}, 972–977 (2016).

\bibitem{Pacheco} 
V. Pacheco-Peña, N. Engheta, Antireflection temporal coatings. {\itshape Optica} \textbf{7}, 323-331 (2020).

\bibitem{PendryRev} 
E. Galiffi \textit{et al.}, Photonics of time-varying media. {\itshape Advanced Photonics} \textbf{4}, 014002 (2022).

\bibitem{Sajjad-review} 
G. Ptitcyn, M. S. Mirmoosa, A. Sotoodehfar, S. A. Tretyakov, Tutorial on basics of time-varying electromagnetic systems and circuits. {\itshape arXiv:2211.13054}, (2022).

\bibitem{Kalluri92} 
D. K. Kalluri, V. R. Goteti, Frequency shifting of electromagnetic
radiation by sudden creation of a plasma
slab. {\itshape Journal of Applied Physics} \textbf{72}, 4575 (1992).

\bibitem{Kalluri93} 
D. K. Kalluri, Frequency shifting using magnetoplasma
medium: flash ionization. {\itshape IEEE Transactions on Plasma Science} \textbf{21}, 77-81 (1993).

\bibitem{Kalluri2010}
D. K. Kalluri, Electromagnetics of time varying complex media: frequency and polarization transformer, (CRC Press, Boca Raton, 2010).

\bibitem{Huanan}
H. Li, S. Yin, A. Alù, Nonreciprocity and faraday rotation at time interfaces. {\itshape Phys. Rev. Lett.} \textbf{128}, 173901 (2022).



\bibitem{temporalaiming}
V. Pacheco-Peña, N. Engheta, Temporal aiming. {\itshape Light Sci Appl,} \textbf{9}, 129 (2020).

\bibitem{Akbarzadeh}
A. Akbarzadeh, N. Chamanara, C. Caloz, Inverse prism based on temporal discontinuity \& spatial dispersion. {\itshape Opt. Lett.} \textbf{43}, 3297-3300 (2018).


\bibitem{Huidobro}
P. A. Huidobro \textit{et al.}, Fresnel drag in space–time-modulated metamaterials. {\itshape Proc. Natl. Acad. Sci.} \textbf{116}(50), 24943–24948 (2019).

\bibitem{Huidobro2}
P. A. Huidobro \textit{et al.}, Homogenization theory of space-time metamaterials. {\itshape Phys. Rev. Appl.} \textbf{16}(1), 014044 (2021).

\bibitem{Fortuno}
K. Bliokh, F. Rodríguez-Fortuño, F. Nori \textit{et al.}, Spin–orbit interactions of light. {\itshape Nature Photonics} \textbf{9}, 796–808 (2015). 

\bibitem{Aiello}
A. Aiello, P. Banzer, M. Neugebauer \textit{et al.}, From transverse angular momentum to photonic wheels. {\itshape Nature Photonics} \textbf{9}, 789–795 (2015). 

\bibitem{Shi}
P. Shi \textit{et al.}, Transverse spin dynamics in structured electromagnetic guided waves. {\itshape Proc. Natl. Acad. Sci.} \textbf{118}(6), e2018816118
(2021).

\bibitem{Eismann}
J. S. Eismann, L. H. Nicholls, D. J. Roth \textit{et al.}, Transverse spinning of unpolarized light. {\itshape Nature Photonics} \textbf{15}, 156–161 (2021). 

\bibitem{Chen}
P. Chen \textit{et al.}, Chiral coupling of valley excitons \& light through photonic spin–orbit interactions. {\itshape Adv. Funct. Mater.} \textbf{8}, 5 (2019). 

\bibitem{Higo}
T. Higo, H. Man, D. B. Gopman \textit{et al.}, Large magneto-optical Kerr effect \& imaging of magnetic octupole domains in an antiferromagnetic metal. {\itshape Nature Photonics} \textbf{12}, 73–78 (2018).

\bibitem{spin}
M. Onoda, S. Murakami, S. Nagaosa, Hall effect of light. {\itshape Phys. Rev. Lett.} \textbf{93}, 083901 (2004).

\bibitem{Bliokh}
K. Bliokh, P. Bliokh, Topological spin transport of photons: the optical Magnus effect \& Berry phase. {\itshape Physics Letters A.} \textbf{333}, 3–4 (2004).

\bibitem{Hosten}
O. Hosten, P. Kwiat, Observation of the spin Hall effect of light via weak measurements. {\itshape Science} \textbf{319}, 5864 (2008).

\bibitem{Zhou}
X. Zhou, Z. Xiao, H. Luo, S. Wen, Experimental observation of the spin
Hall effect of light on a nanometal film via weak measurements. {\itshape Phys. Rev. A} \textbf{85}, 043809 (2012).

\bibitem{Ari}
I. V. Lindell, A. H. Sihvola, S. A. Tretyakov, A. J. Viitanen, Electromagnetic Waves in Chiral \& Bi-isotropic Media. {\itshape Norwood, MA: Artech House}, (1994).

\bibitem{CONDON}
E. U. Condon, W. Altar, H. Eyring, One‐electron rotatory power. {\itshape J. Chem. Phys.} \textbf{5}, 753 (1937).

\bibitem{Morgen}
F. Morgenthaler, Velocity modulation of electromagnetic waves. {\itshape IEEE Trans. Microw. Theory Tech.} \textbf{6}, 167 (1958).

\bibitem{Serdyukov}
A. N. Serdyukov, I. V. Semchenko,  S. A. Tretyakov, A. H. Sihvola, Electromagnetics of Bi-anisotropic Materials: Theory and Applications, {\itshape Amsterdam: Gordon and Breach Science Publishers}, (2001).

\bibitem{JEWA}
S. T. Tretyakov, F. Mariotte, Maxwell Garnett modeling of uniaxial chiral composites with bianisotropic inclusions, {\itshape Journal of Electromagnetic Waves and Applications},  \textbf{9}, 1011-1025, (1995).




\end{thebibliography}

\section*{Acknowledgments}
This work was supported by the Academy of Finland under grant 330260.

\section*{Author contributions}
M.S.M conceived the original idea, M.H.M and S.A.T helped further formulating the concept. M.H.M performed the theoretical analysis with input from M.S.M and S.A.T. M.H.M performed the numerical analysis, and wrote the manuscript. M.S.M and S.A.T reviewed and edited the manuscript. S.A.T supervised the work.

\section*{Competing interests}
All authors have no competing interests

\section*{Data and materials availability}
All data is available in the manuscript or the supplementary materials.

\section*{Supplementary materials}
Materials and Methods\\
Supplementary Notes 1 to 4\\
Figs. S1 and S2\\
References \textit{(30-34)}

\end{document}